# ULTRALIGHT SOLAR POWERED HYBRID RESEARCH DRONE

Cs. Singer, German Aerospace Center (DLR), Institute of Solar Research, Pfaffenwaldring 38-40, 70569 Stuttgart
csaba.singer@dlr.de

**Introduction:** This proposal concerns an ultralight solar powered research drone, which is VTOL capable and which absolves its cruise flight in an efficient axis symmetric configuration. It is ideal for the over ground exploration of Mars, as it can carry divers sensor systems, while the structural mass is minimized due to a lighter than $CO_2$ ($LTCO_2$, on earth lighter than air) concept, combined with a lightweight construction. The buoyant $LTCO_2$ gas can be among others air, $N_2$, $O_2$, He or $H_2$. The structure stability of the ultra lightweight drone will be achieved with a circular filigree fiber composite ring and a slightly pressurized gas cell made from ultra dense foil. Synergetic advantages comprise the facts, that every technical component, which is necessary for a rotation symmetric VTOL, also can be used for a mirror symmetric cruise, whilst during take-off, cruise-flight and landing the demanded lift will be generated in a hybrid way using dynamic lift and static buoyancy. The power supply consists of a combination of batteries and photovoltaic cells and optional regenerative fuel cells. The lenticular shell is optimally designed to accommodate a maximized photovoltaic surface at maximal $LTCO_2$ gas volume and at minimized structural mass.

**Functionality:** For the explanation of the functionality Fig.1 to Fig.3 and an imaginary flight from a location A to a location B is used.

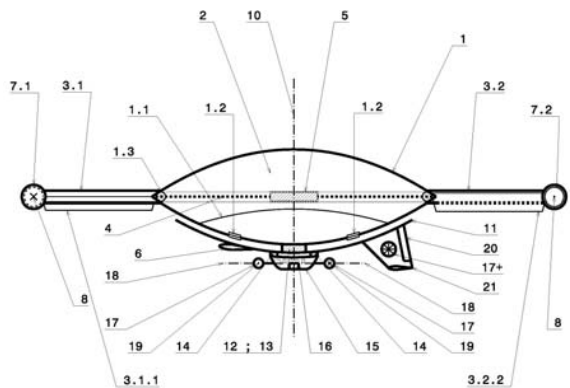

**Fig. 1** Rotation symmetric arrangement (side view)

The research drone in Fig. 1 passed the entry phase an reached the Mars surface. After deconvolution (e.g. like Pathfinder) it loaded the batteries and waits ready for takeoff at position A for an autonomous or manual command. The first command is to takeoff vertically and to reach a given altitude. The lateral mounted full symmetric wings with flapped tabs (shown in Fig. 2), the engines at the end of the wings and the lenticular hull are configured rotation-symmetrically.

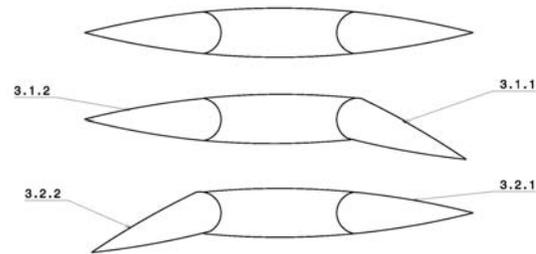

**Fig. 2** Fully symmetric wings with inclinable tabs

Wings, engines and hull are the lifting components of the entire system. These are separated with a bearing from the carried components, which are the support for the payload, the payload itself and the energy support (e.g. batteries). While takeoff trust from the engines induces a rotation of the lifting components and dynamic lift will be generated by the flapped tabs. After takeoff the proposed aircraft climbs to a required altitude that is below the altitude, at which the buoyant gas fills out the entire hull volume as the gas will expand with increased altitude.

For purposes of clarity Fig. 1 and Fig. 3 contain components which are optional (o) for planetary exploration, for example if the rotation of the payload has to be avoided for some reason. The following list the technical components.

| | | | |
|---|---|---|---|
| 1 | lenticular hull | 9 | direction of zero lift |
| 1.1 | gas-cell-construction | 10 | axis of rotation |
| 1.2 | blower | 11 | lightweights rail (o) |
| 1.3 | carbon-fibre-ring | 12 | radial bearing (o) |
| 2 | volume of hull | 13 | sliding contact (o) |
| 3.1 | wing 1 | 14 | gliding bearing (o) |
| 3.2 | wing 2 | 15 | cabin or payload |
| 3.1.1 | front tab | 16 | power supply |
| 3.2.2 | back tab | 17 | optional engine (o) |
| 4 | axis of rotation (wing) | 18 | axis of rotation (o) |
| 5 | rotatable mechanism (o) | 19 | thrust vector |
| 6 | canard | 20 | vertical tail |
| 7.1 | main engine 1 | 21 | horizontal tail |
| 7.2 | main engine 2 | 22 | horizontal plane |
| 8 | thrust vector | 23 | emerg. parachute (o) |

If a new command was given to fly from A to B the engines stop and the rotation dies out. The research drone loses altitude slowly, while the tabs are moved into the neutral position by the on board control. When the rotation died out the on board control moves the tabs in a way that he whole drone is configured mirror symmetrically. By giving parallel thrust on both sides and moving the tabs for a dynamic lift, the proposed aircraft flies forward in the mirror symmetrical configuration as shown in Fig. 3. Induced dynamic lift and additional buoyancy are sufficient to keep the required altitude. While cruse flight the payload can be moved along its support into the required position to reach dynamic stability. The controlled horizontal tail and canard care for the equilibrium while cruse flight, whereas the approach angle of the hull is hold around zero. Finally the thrust of the engines can be decreased, when point B is near. The configuration will be changed again to be rotationally symmetrical, when the speed is reduced and a precision vertical landing should be carried out. While flight but also on ground the sensor equipment on board collects significant data related to the discovery of our neighbor planet.

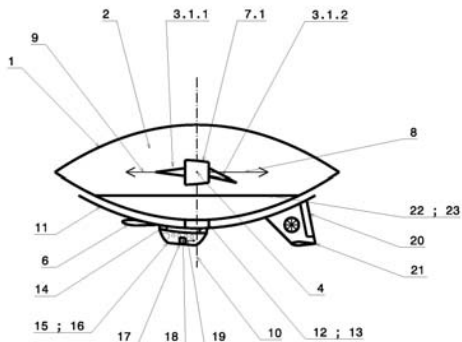

**Fig. 3** Mirror symmetric arrangement (side view)

**Innovative Aspects:** The proposed research drone combines three approved main concepts of human flight. It needs no infrastructure for takeoff and landing and is able to reach a higher cruising speed compared to airships and helicopter. It is built out of lightweight high-tech materials like fiber composites, sailcloth and high dense foils, while only a low amount of metal is applied to the structure. It uses solar energy as a regenerative energy source, whose conversion is accomplished by solar cells mounted at the upper surface of the hull and batteries and/or regenerative fuel cells. It allows autonomous non-stop flight if required. In the case of emergency the construction behaves like a parachute, which competes with the security strategy of all conventional flying concepts.

**Development:** After the idea was applied for a patent (DE102006028885A1) the feasibility study was carried out with a simplified experimental approach, which showed that it is possible to carry out the basic abilities, VTOL, mirror symmetric cruse flight and parachute characteristic with one and the same equipment like proposed. After this a basic modeling and design tool was programmed, which helps to estimate the dimensions, power supply and the reachable performance. No validation of this tool took place yet.

**Additional Information:** Lighter than air constructions with lenticular hulls that reached the development stage of prototyping were the models XEM-1 to XEM-4 from LTAS/CAMBOT LLC, remotely piloted lenticular airships, which were built from 1974 until 1981 as a demonstrator and for filming, video observation and telecommunications work and their three full scaled rigid airship variants: SPATIAL-MLA-24, MLA-32-A and MLA-32-B [1,2]. Also ALA-600 Thermoplane, an airship filled with both helium and hot gas, which was designed to operate with heavy loads, without a base or mooring mast was finalised in 1989, whereas ALA-40-01 ground tests started in 1992 [2]. Actual efforts to realize airships with lenticular hulls are the Alizé concept, AirFerry and AeroRaft [3,4,5]. Also SLTA is assessing unmanned lenticular airship configurations for cargo applications [6]. LTA references can be found in [7].

**References:**

[1] JANE's, "All the World's Aircraft" Coulsdon, Surrey UK; Alexandria VA: Jane's Information Group. Annual
[2] NAYLER, A., "Airship development world wide - A 2001 Review", Airship Association Ltd. London, 2001
[3] BALASKOVIC, P., "Presentation d'un Aeronef de conception entierement nouvelle: L'Alizé", LTA Corporation Trust Center, 2007
[4] KÜNKLER, H., "A Hybrid Aircraft for Low-Infrastructure General Transport", Intellect. Property, 2007
[5] LUFFMAN, C.R., "AeroRaft™ - A Brief Insight", LTA Solutions, 2005
[6] BOCK, J.K., "Autonomous Cargo Airships Operations System", SLTA, 2008
[7] KHOURY G.A., GILLET J.D., „Airship Technology", Cambridge University Press, ISBN-13: 9780521607537, 1999